\documentclass[format=acmsmall, review=false, screen=true, authorversion=true]{acmart}

\usepackage[utf8]{inputenc}

\usepackage{booktabs} 

\usepackage[ruled]{algorithm2e} 

\SetAlFnt{\small}
\SetAlCapFnt{\small}
\SetAlCapNameFnt{\small}
\SetAlCapHSkip{0pt}
\IncMargin{-\parindent}

\usepackage{amssymb}
\usepackage{url}

\usepackage{tabularx}
\newcolumntype{Y}{>{\centering\arraybackslash}X}

\usepackage{amsmath}
\usepackage{commath}
\usepackage{multirow}
\usepackage{hhline}
\usepackage{rotating}

\newcommand{\MAXATF}[0]{\mathrm{MAX\_ATF}}
\newcommand{\trendfactor}[1]{\mathit{trend\_factor_{#1}}}
\newcommand{\atf}[1]{\mathit{atf_{#1}}}
\newcommand{\adjustedatf}[1]{\mathit{adj\_atf_{#1}}}
\newcommand{\expecttrust}[1]{\mathit{expect\_trust_{#1}}}
\newcommand{\currenttrust}[1]{\mathit{current\_trust_{#1}}}
\newcommand{\aggregatetrust}[1]{\mathit{aggregate\_trust_{#1}}}
\newcommand{\trustvalue}[1]{\mathit{trust\_value_{#1}}}
\newcommand{\threshold}[0]{\mathit{threshold}}
\newcommand{\changerate}[1]{\mathit{change\_rate_{#1}}}
\newcommand{\sendproportion}[1]{\mathit{send\_proportion_{#1}}}
\newcommand{\sendingamount}[1]{\mathit{sending\_amount_{#1}}}
\newcommand{\maxsendingamount}[1]{\mathit{maximum\_sending\_amount_{#1}}}

\usepackage{subcaption}
\usepackage{graphicx}
\graphicspath{{img/}}

\acmJournal{TOIT}
\acmVolume{19}
\acmNumber{4}
\acmArticle{46}
\acmYear{2019}
\acmMonth{9}
\acmDOI{https://doi.org/10.1145/3329250}

\begin{document}

\title{The influence of trust score on cooperative behavior}

\author{Claudia-Lavinia Ignat}

\affiliation{
  \institution{Université de Lorraine, CNRS, Inria, LORIA}
  \postcode{Inria Nancy – Grand-Est, 615 rue du Jardin Botanique, 54600}
  \city{Villers-lès-Nancy}
  \country{France}}
  \email{claudia.ignat@inria.fr}
  
\author{Quang-Vinh Dang}
\authornote{now at Data Innovation Lab, Faculty of Information Technology, Industrial University of Ho Chi Minh City, Vietnam}
\affiliation{
  \institution{Université de Lorraine, CNRS, Inria, LORIA}
  \postcode{Inria Nancy – Grand-Est, 615 rue du Jardin Botanique, 54600}
  \city{Villers-lès-Nancy}
  \country{France}}
  \email{quang-vinh.dang@inria.fr}
\author{Valerie L. Shalin}
\affiliation{
  \institution{Department of Psychology and Kno.e.sis, Wright State University}
  \postcode{Fawcett Hall 447, 3640 Colonel Glenn Hwy, Dayton, OH 45435-0001}
  \country{USA}}
\email{valerie.shalin@wright.edu}


\begin{abstract}

The assessment of trust between users is essential for collaboration. General reputation and ID mechanisms may support users' trust assessment. However, these mechanisms lack sensitivity to pairwise interactions and specific experience such as betrayal over time.  Moreover, they place an interpretation burden that does not scale to dynamic, large-scale systems.
While several pairwise trust mechanisms have been proposed, no empirical research examines trust score influence on participant behavior. We study the influence of showing a partner trust score and/or ID on participants behavior in a small-group collaborative laboratory experiment based on the trust game. We show that trust score availability has the same effect as an ID to improve cooperation as measured by sending behavior and receiver response. Excellent models based on the trust score predict sender behavior, and document participant sensitivity to the provision of partner information.  Models based on the trust score for recipient behavior have some predictive ability regarding trustworthiness, but suggest the need for more complex functions relating experience to participant response. We conclude that the parameters of a trust score, including pairwise interactions and betrayal influence the different roles of participants in the trust game differently, but complement traditional ID and have the advantage of scalability.

\end{abstract}

%
%



\maketitle

\renewcommand{\shortauthors}{C.-L. Ignat et al.}

\section{Introduction}

Flattened organizational hierarchies promote reliance on direct peer-to-peer interactions \cite{mcchrystal2015team}.
However, this increases both the number of interactions and critically, the number of peers interacting with each other.
At the same time, \emph{ad-hoc} work groups increasingly respond to transient need, as in  Wikipedia modifications, crisis response, political activism and software development. 
Technology facilitates these ad-hoc work groups, allowing users from different locations to collaborate without the need for face-to-face interaction.

However, psychological considerations exist concerning globally distributed work, particularly the cognitive demand associated with developing, maintaining and accessing a large number of interactions with a large network of partners. \citeN{rentsch2001great} identify shared knowledge, including schemas, goals and values as fundamental to effective collaboration.  Accordingly, overall performance benefits from participants who share expertise and world views. Critically, acquiring knowledge about a collaborator's expertise and worldview is a \emph{process}. Some researchers specifically identify \textit{transactive memory} as essential to successful performance in distributed work \cite{DBLP:conf/pacis/Chang04}, allowing participants to solicit assistance and information from the best resource for the task at hand. Exploiting transactive memory entails cognitive demand for encoding, retrieving and updating representations of each participant's capabilities, stored in declarative memory. 

 In a conventional work environment, participant names serve as a retrieval cue for recalling a participant's expertise and personally observed, previous behavior maintained in declarative memory. Consistent with this practice, virtual identities or IDs, either assigned by a system or chosen by users, distinguish between participants. Similar to the notion of branding, when participant A sees the ID of B, participant A can recall the experiences she had with B, and engage accordingly.
 Of course, nonsense ID strings such as ``p67718an22187bOz" \cite{dix2009human} are not remembered well. More concerning is that  psychological research has established the persisting response time penalties of increasing the size and interconnectedness of declarative content such as ID \cite{anderson1999fan}.  As a result, increasingly large, dense networks with rarely accessed nodes, such as those made possible by internet collaboration, pose retrieval problems, and hence access to the knowledge that supports effective collaboration. 
 
 In this paper we seek a computationally derived, behavior-based substitute for the above-mentioned demands that scales better than ID with increasing network size.
 We suggest the presentation of trust scores as a substitute for maintaining detailed, qualitative accounts of prior experience between partners. We use the definition of trust as ``a cognitive learning process obtained from social experiences based on the consequences of trusting behaviors'' \cite{ChoCA15}, where trust is built based on observations in the past.

\textit{Trust} and \textit{reputation} are sometimes used interchangeably \cite{vu2010trust,DBLP:journals/cn/Pecori16}. Though related, they are not the same constructs \cite{fetchenhauer2009people}. Consistent with \cite{breitmoser2015cooperation}, we consider reputation as the \emph{ collective opinion of a community} regarding a particular participant, while trust is the \emph{specific} relationship between a pair of participants.  Participant reputation is a \textit{global} value, while trust in a participant is a \textit{personal} value and differs by partners \cite{DBLP:journals/aamas/HoelzR15}. This distinction allows us to accommodate the different concerns and corresponding weights that one participant has relative to another.  Psychological research supports the claim that different participants view the trustworthiness of the same target differently \cite{bergman2010asymmetry}. Personality and perceptual bias may also influence an observer's assessment of a target's trustworthiness. For these reasons we seek a metric that characterizes pairwise trust.   

Using these definitions,  the widely used Internet scoring systems are reputation--not trust--systems \cite{resnick2000reputation}. Examples include the Amazon reputation score, calculated by averaging all rating scores from \emph{all buyers}, such that every buyer will see the same score when they examine the seller's profile. The heart of a reputation system is therefore indirectness: one benefits from interacting with participants who have been shown to be trustworthy with other people.   Studies such as \cite{tadelis1998s} established a strong connection between reputation and name. Bolton \cite{bolton2002effective} demonstrated the effectiveness of reputation scores in e-commerce.  General reputation is particularly relevant in the absence of repeated interaction with a particular individual. In \cite{Resnick2006} the authors conducted a controlled field experiment of an Internet reputation mechanism where they ruled out several potential confounds appearing in previous observational studies such as seller skill, product quality and seller responsiveness to customer inquiries. The study found that sellers with high reputation fared better; buyers were willing to pay a well established reputation
seller 8.1\% more on average than a new seller for the same item. However, the study did not analyze how repeated interaction between a customer and a seller is influenced by reputation score.

The main limitation of reputation systems is their vulnerability to third party manipulation \cite{DBLP:journals/csur/HoffmanZN09}. Suppose Alice wants to know the reputation score of Bob, that is, the community opinion of Bob. Alice may query other users about Bob, say Carol or Dave. Or she may acquire a reputation score from a central server, which has collected the opinions about Bob from all users. In any case, Alice relies on information from third parties. The information might not be available, i.e. a central server might go down or is unavailable in a peer-to-peer system, or Carol never answers Alice. The information might not be reliable, i.e. Carol might give an unfair opinion of Bob \cite{JosangDSS07}. Additionally Bob can create multiple virtual identities to  provide deceptively high rating scores for himself. In order to address reputation attacks some researchers  \cite{DBLP:conf/chi/SangerHGBLS16} have proposed an enhanced presentation of reputation data.  Their interactive visualization  increased a participant's ability to detect and understand malicious seller behavior in e-commerce. While the approach was efficient for participants with less experience, the additional information also distracted some participants.

Reputation systems are also vulnerable to the \textit{playbook} attack \cite{josang2009challenges} where a service provider provides bad service only to a subset of participants, gaining both revenue and reputation score at the expense of a few unhappy participants. In this case, the assumption that a participant behaves identically with all participants is clearly wrong. Nevertheless, in the popular averaging reputation system \cite{JosangDSS07}, all behaviors have the same weight regardless of the surrounding context. There is not yet an effective technique to deal with the playbook strategy \cite{DBLP:journals/imds/SunK14}.  More generally, the psychometric foundations of reputation scores are unclear, such as the treatment of variability or the weighting of recent information, especially betrayal. 

Furthermore, reputation scores lack personalization. According to \citeN{DBLP:conf/icdcsw/WangV07} a personalized score is required in the  presence of subjective factors, i.e. user needs or interests.  A \emph{personal trust} scoring system can accommodate subjectivity. Such a trust score is ideally calculated and attached to a participant. Because the trust score reflects personal experience between pairs of users, the playbook attack is not possible. A user can compute the trust score of a partner locally without querying information from third parties. Crucially, with effective trust scoring, participants do not need to recall anything. Continued, context sensitive interaction therefore proceeds with limited cognitive demand.

We employed a dynamic trust function that calculates participant trust values based on behavioral history \cite{DBLP:conf/trustcom/DangI16}. 
We examined the effect of presenting this trust metric and ID on trust behavior, with a post-hoc analysis of reputation as a predictor. For this purpose, we adapted the \textit{trust game} \cite{berg1995trust}, a money exchange game that is widely used in economics to study human trust behavior \cite{johnson2011trust,lewicki2015trust}. We particularly investigated whether the availability of either trust scores or ID improves user cooperation. The set of analyses and results of our study are available at \url{https://github.com/coast-team/trust_influence_analysis}.

\section{Trust Game}

We employed the trust game as an analogue for the exchanges between pairs of interdependent participants in distributed work, where partner identity is relevant to the assessment of partner behavior. Numerous studies documented behavior in the trust game context.  This provides experimental design and performance standards and allows us to associate observed behavior with our specific manipulations and/or dismiss idiosyncrasies as non-influential. In this section we describe the trust game and give an overview of several studies of cooperation that employed the trust game.

\begin{figure}[t]
    \centering
    \includegraphics[width=0.5\textwidth,trim={0 .5cm 0 1.2cm},clip]{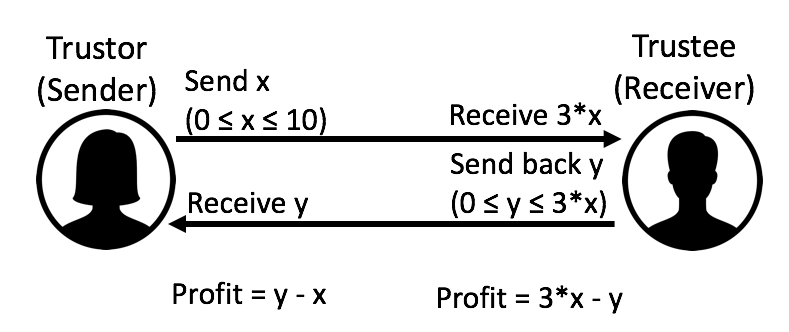}
    \caption{One trial trust game.}
    \label{fig:trust_game}
\end{figure}

\citeN{berg1995trust} developed the trust game, or the ``investment game",  to study economic reciprocity. Participants are organized in pairs. For each pair, one participant is assigned the role of ``sender" while the other is assigned the role of ``receiver". The two roles are sometimes called ``trustor" and "trustee" respectively.  As shown in Figure \ref{fig:trust_game}, initially the sender sends an integer amount between 0 and 10 units  to the receiver. The receiver gains three times the amount sent. For instance, if the sender sent $7$ money units, the receiver will gain $3 * 7 = 21$ units. Subsequently, the receiver can select an amount between 0 and the gained amount (in this case, $21$) to return to the sender. However, the returned amount is not further multiplied. Suppose the receiver returned $11$. The final payoff to the sender is $11$ units, and the payoff to the receiver is $21 - 11 = 10$ units. 

The game pits joint payoff against individual payoff. Joint payoff is maximized if the sender sends $10$ to the receiver, so the total profit is $20$, calculated as the total payoff of $30$ minus the initial endowment of $10$ to the sender. However, assuming that participants only seek to maximize their own profit, normative game theory predicts that the sender will send $0$ and upon receiving any sum the receiver will send back $0$. Any amount other than 0 will reduce the receiver's profit. According to normative theory for the one-round trust game, the sender knows this fact, so at her turn, she should send 0 to the receiver. If she sends any greater amount, she must assume that she will not receive anything back \cite{camerer2003behavioral}.  

In fact, participants do not behave according to normative theory, but choose to maximize their joint profit. The trust game is therefore considered to be \textit{cooperative} \cite{cesarini2008heritability,balliet2013trust}, or said differently, increases in payoff reflect cooperation. Researchers interpret the initial sending amount as an expression of sender trust in the receiver \cite{glaeser2000measuring}. Receiver response is an expression of trustworthiness \cite{fehr2003nation}. \citeN{brulhart2012does} equate trust game behavior as a measure of trust and trustworthiness. 

Repetition of the exchange enhances cooperation \cite{cochard2004trusting}. 
In the repeated game, trust potentially accrues as both players react to their partner's past behavior in subsequent rounds.  To our knowledge, the theoretical analysis for participant behavior in the infinitely repeated trust game is still an open question \cite{bruttel2012infinity,breitmoser2015cooperation}. However, we note that the two players experience consequence at different times.  The sender receives immediate feedback regarding his trust in the form of the receiver's response.  The receiver on the other hand will not incur consequence to his trustworthiness until a subsequent exchange with this sender. Studies of the repeated trust game \cite{engle2004evolution} document a decline in cooperation towards the end of a session with known length.   

Some research work examines the influence of reputation in the trust game or public good games. \citeN{semmann2004strategic,DBLP:journals/geb/HuckLT12} associate reputation with user identity, showing  that cooperation declines when individual identities
switch from being recognizable to being unrecognizable.  \citeN{DBLP:conf/hci/BenteDRRL14} tested the influence of avatar and reputation levels on buyer decisions, i.e. senders in the trust game. Both reputation scores and avatars can encourage the investment decision of buyers. However, the authors did not study the behavior of sellers, i.e. receivers in the trust game. Moreover, reputation scores were artificial rather than computed from real behavior. \citeN{keser2003trust} represents user reputation as a set of trinary ratings ("positive", "negative" or "neutral") manually assigned by senders to rate the behavior of receivers in the trust game. These ratings were presented to subsequent senders before they made their decisions. The study found that reputation information  significantly increases the overall cooperation levels of the game. A follow-up experiment \cite{boero2008reputation} employed three games: in the first one, similar to \citeN{keser2003trust}, senders could rate receivers; in the second game receivers could rate senders and in the third one both senders and receivers could rate their partners. The study confirmed the findings of \citeN{keser2003trust} regarding senders and reported on similar results regarding receivers. The study found that a bidirectional reputation scheme does not perform necessarily better than a single-way reputation scheme. However, in both studies \citeN{keser2003trust} and \citeN{boero2008reputation} reputation levels were manually assigned by participants, who had to examine all previous partner reputation levels before making a decision.  

Our view is that general reputation is not always available, and that a participant is only aware of his own experience with specific partners but is not aware of, or does not necessarily want to rely on, other users' experience with those partners.
In particular, we examined the influence of an available \emph{trust score} on participant behavior, controlling for the availability of user ID. The trust score is automatically computed by the system based on participants behavior during the game, taking variability and misbehavior into account. There is no burden on participants to assign or calculate partner trust scores manually.  We study the influence of trust score for both sender and receiver roles. We did depart from the standard repeated trust game paradigm in one notable way: to maximize power, our manipulations concerning the availability of trust score or participant name 
are within-subjects.

\section{Preliminary study}

We present a preliminary analysis of the predictive power of participants' future behavior in the trust game comparing trust and reputation scores.

According to \citeN{DBLP:journals/ecr/Malaga01}, reputation score (and by inference) trust score comprises a prediction about future behavior. For instance, if Alice has a high score, we could expect that she will behave well in the future. If she fails to do so, the score assigned to her is inaccurate. Below we compare the relative predictive power of trust and reputation scores.  

We employed two external datasets from two repeated simple trust game experiments independently conducted by \citeN{dubois2012does} and \citeN{bravo2012trust}. The experiment in \citeN{bravo2012trust} involved 108 participants and contained five rounds. The experiment in \citeN{dubois2012does} involved 36 participants and contained ten rounds. Both experiments employ groups of 18 participants. For computing trust scores we employed the trust function proposed in \citeN{DBLP:conf/trustcom/DangI16}, shown to reflect and predict participants' behavior in the repeated trust game, with resistance  to  fluctuating  participant  behavior. As a reputation measure we used participants' average sending proportion up to the moment reputation is computed, which is similar to many real-world reputation scoring methods \cite{JosangDSS07,DBLP:journals/jcm/TavakolifardA12}.

We conducted one regression analysis using the trust score computed by our trust function as a predictor and with observed sending proportion as the criterion. Starting with round 4 when the trust metric has stabilized, we predicted the send proportion of participants using the trust score calculated after the previous round.  We employed a similar regression analysis with reputation score as a predictor and sending proportion as the criterion. The results for the sender role appear in Table \ref{tab:regression_external_sender} and for the receiver, in Table \ref{tab:regression_external_receiver}. The corresponding t-values for the trust value assigned to each participant in predicting their future behavior are all significant for both senders and receivers, i.e. the trust score calculated by our trust function is predictive for external datasets. Moreover, adjusted $R^2$ values are higher for predictive models using trust values than for reputation values in all cases except for round 7 and 8 for senders in the Dubois dataset and equal in round 4 of the Bravo dataset for receivers.

\begin{table}[t]
    \small
    \centering
    \begin{tabular}{@{}lrrccc@{}}
        \toprule
        \bf Dataset  & \bf df & \bf t-value & \bf Adj. $R^2$ & \bf t-value & \bf Adj. $R^2$\\
        \bf          & \bf    & \bf  for trust & \bf for trust & \bf  for reputation & \bf for reputation\\
        \midrule
        Bravo dataset (round 4)    & 106 & 7.85*** & 0.36 & 3.05** & 0.19 \\
        Bravo dataset (round 5)    & 106 & 10.0*** & 0.48 & 8.86*** & 0.42 \\
        Dubois dataset (round 4)   & 34 &  4.41*** & 0.35 & 3.24** & 0.21\\
        Dubois dataset (round 5)   & 34 & 4.51*** & 0.36 & 2.84** & 0.17\\
        Dubois dataset (round 6)   & 34 & 4.68*** & 0.37 & 4.26*** & 0.32\\
        \textit{Dubois dataset (round 7)} & \textit{34} & \textit{4.05***} & \textit{0.31} & \textit{4.29***} & \textit{0.33}\\
        \textit{Dubois dataset (round 8)}   & \textit{34} & \textit{4.15***} & \textit{0.32} & \textit{4.83***} & \textit{0.39}\\
        Dubois dataset (round 9)   & 34 & 4.25*** & 0.33 & 3.17** & 0.21\\
        Dubois dataset (round 10)  & 34 & 4.52*** & 0.36 & 2.36* & 0.11\\
        \bottomrule
    \end{tabular}
    \caption{Regression analysis of our trust function and reputation applied on external datasets for sender role. Italicized entries have a higher or equal Adj. $R^2$ for reputation than for trust.
     `*'$\emph{p} < 0.05$,  `**' $\emph{p} < 0.01$, `***' $\emph{p} < 0.001$.}
    \label{tab:regression_external_sender}
\end{table}

\begin{table}[b]
    \small
    \centering
    \begin{tabular}{@{}lrrccc@{}}
        \toprule
        \bf Dataset  & \bf df & \bf t-value & \bf Adj. $R^2$ & \bf t-value & \bf Adj. $R^2$\\
        \bf          & \bf    & \bf  for trust & \bf for trust & \bf  for reputation & \bf for reputation\\
        \midrule
        \textit{Bravo dataset (round 4)}    & \textit{93} & \textit{4.72***} & \textit{0.18} & \textit{4.71***} & \textit{0.18}\\
        Bravo dataset (round 5)    & 64 & 5.04*** & 0.27 & 4.61*** & 0.24\\
        Dubois dataset (round 4)   & 30 & 3.84*** & 0.31 & 3.15** & 0.22\\
        Dubois dataset (round 5)   & 31 & 4.58*** & 0.35 & 2.95** & 0.19\\
        Dubois dataset (round 6)   & 31 & 6.06*** & 0.53 & 2.20* & 0.11\\
        Dubois dataset (round 7)   & 29 & 6.52*** & 0.58 & 2.93** & 0.20\\
        Dubois dataset (round 8)   & 30 & 6.69*** & 0.64 & 4.88*** & 0.42\\
        Dubois dataset (round 9)   & 26 & 3.86*** & 0.34 & 1.59 & 0.05\\
        Dubois dataset (round 10)  & 27 & 4.88*** & 0.45 & 4.38*** & 0.39\\
        \bottomrule
    \end{tabular}
    \caption{Regression analysis of our trust function and reputation applied on external datasets for receiver role. Italicized entries have a higher or equal Adj. $R^2$ for reputation than for trust. `*'$\emph{p} < 0.05$,  `**' $\emph{p} < 0.01$, `***' $\emph{p} < 0.001$.}
    \label{tab:regression_external_receiver}
\end{table}

This preliminary analysis provides compelling evidence for the predictive power of trust scores. The trust model requires \emph{less raw information} (only information observed by the user) than the reputation model, which requires complete information from all users. These are somewhat surprising findings given that none of these participants were aware of their partners. In fact, the trust function uses less raw data but has more contextual parameters than reputation, accounting for partner, cumulative behavior over time, and punishment of misbehavior. In the next sections we present our research questions and experimental design for demonstrating the influence of trust scores on user cooperative behavior and how our trust metric predicts behavior.

\section{Research questions}

We study how the availability of partner trust score and ID impacts participant behavior and the appropriateness of the used trust metric for computing trust scores in the repeated trust game. We grouped our research questions as follows:

\textbf{RQ1} \textit{Does showing partner trust score or ID change user cooperative behavior?}
If so, is there a significant difference in cooperative user behavior with only  trust scores relative to ID only? Is there a significant difference in user cooperative behavior resulting from the availability of both trust score and ID compared to the availability of only one of these two features? Does cooperative behavior change over time?

\textbf{RQ2} \textit{Does the trust calculation predict participant's future behavior?}
\textit{Do participants follow the guidance of the trust calculation?}

As senders and receivers have two different roles and may behave differently, we analyze these research questions separately from both the senders' and receivers' points of view.

\section{Methods}

\begin{table}[b]
    \small
    \centering
    \begin{tabular}{@{}>{\centering\arraybackslash}m{0.3cm} >{\centering\arraybackslash}m{.3cm} >{\arraybackslash}p{5.9cm} >{\arraybackslash}p{5.9cm}@{}}
        \toprule
         & & \multicolumn{2}{c}{\bf ID presented} \\

        && \multicolumn{1}{c}{\bf False} & \multicolumn{1}{c}{\bf True} \\
        \cmidrule{3-4} \\[-3em]
        \multirow{2}{*}{\rotatebox[origin=c]{90}{\parbox{2.3cm}{\bf Trust presented}}} & \rotatebox[origin=c]{90}{\parbox{1.5cm}{\bf False}} & {\bf Simple Game}: The trust game when participants are given no information about partners & {\bf Identity Game}: The trust game when participants are given only partner ID
 \\[-.8em]
        & \rotatebox[origin=c]{90}{\parbox{1.3cm}{\bf True}} & {\bf Score Game}: The trust game when participants are given only trust scores of partners & {\bf Combined Game}: The trust game when participants are given both trust the scores and ID of their partners
 \\[0em]
        \bottomrule
    \end{tabular}
    \caption{Game descriptions}
    \label{fig:game_description}
\end{table}

\subsection{Participants}

Participants were recruited through a public announcement. Five independent groups of six participants resulted in a total 30 of participants.  Four of the five groups included one female participant, while the fifth group included two female participants.  The ages of participants ranged from 19 to 45 with an average age of 28.5. 

Typically researchers compensate participants using an exchange rate between virtual money in the experiment and real money, then pay the participants an amount based on how much they earned during the experiment. To assure continuing incentive throughout the session, each person who participated received a coupon of ten euros, but the person who earned most, i.e. who had the highest payoff among other people in the group, received an additional coupon of ten euros. 

\subsection{Task}

In each game a participant played at least $25$ rounds with the other five partners in the group in a random order, namely five rounds with each of these partners where she served as sender and receiver equally often. At the beginning of the first game each participant received $10$ money units. In each round, the sender moved first. She knew how much money she had, and had to decide the amount she wanted to send to the receiver. After that, the receiver received a message indicating  how much she had at the beginning of this round, how much she received from the sender, and how much she will have after having received. Then, the receiver decided how much she wanted to return.

\subsection{Independent Variables}
We crossed the availability of ID and partner trust scores to create four different games as shown in Table \ref{fig:game_description}.  IDs, such as ``Mr. Black" or ``Mrs. Green", were assigned to participants, fixed during a game and varied between games. Trust scores were calculated as in \citeN{DBLP:conf/trustcom/DangI16} (see Appendix \ref{ap:trust_calc}). Trust scores were always calculated for each participant in a pair, but only displayed according to experimental condition and only partner scores were available. The theoretical trust score value ranges from 0.0 to 1.0 inclusive, presented when available with two significant digits.  Participants started with the neutral value of $0.5$ \cite{DBLP:journals/tec/AbbassGP16}.

We calculated user reputation score as distinct from trust score by averaging all previous sending proportion amounts of that user in both roles sender or receiver. 

\subsection{Design}

The experimental conditions were organized as a split-plot factorial with group as a between subjects factor and Show-ID and Show-Trust as within subjects, such that each group of six participants participated in the set of four randomly ordered games. In each round, participants were paired randomly within their group and assigned randomly the sender or receiver role. We ensured that within each game, a participant was paired with a particular other participant at least five times.

\subsection{Dependent Measures}

The four dependent measures used in our study are: \textbf{sending proportion by senders}, \textbf{sending proportion by receivers}, \textbf{average sending proportion by senders} and \textbf{average sending proportion by receivers}.

\textbf{Sending proportion by senders} is the net amount the sender sends to the receiver over 10, which is the maximum amount the sender could send.

\textbf{Sending proportion by receivers} is the net amount the receiver sends back over the amount she received after being tripled. 

 Other studies \cite{burks2003playing,dubois2012does} also used sending proportion measures in order to normalize the sending behavior of receivers for comparison. For example, sender A sent 6 to receiver B, and B sent back 9 to A. In this round, the net sending amount of A and B are 6 and 9 respectively, the sending proportion of A is $6/10=0.6$ and the sending proportion of B is $9/18 = 0.5$.

Consistent with \citeN{burks2003playing,dubois2012does} for all analyses of receiver behavior, we eliminated the zero transaction between the sender and the receivers, (i.e. the sender sends $0$ and the receiver is obliged to send  $0$), for two reasons.  First, receiver behavior is completely determined by the sender, so that the receiver's behavior is not informative. Moreover, in this case, the sending proportion for the receiver ($0$ divided by $0$) is not calculable. We note that the zero-sending amount is retained in the analysis of sender behavior.

For the sender, there are exactly 375 sending proportion data points in each game ($25/2$ senders $\times 6$ players in a group $\times 5$ groups). For receiver, the number of sending proportion data points varies between 250 and 340 due to the elimination of the zero transaction.

\textbf{Average sending proportion by senders} is the average of sending proportions by each sender over all trials in the game.   Taking an average distributes the effect of the zero transaction and also eliminates trial as a repeated factor in analysis.

\textbf{Average sending proportion by receivers} is the average sending proportion the receiver sends back to the sender over all trials in the game, without the zero transaction case.

There are 30 average sending proportion data points corresponding to 30 participants, for both sender and receiver. For the receiver, the zero-transaction data is removed before calculating the means.

\subsection{Procedure} 
All groups participated independently using z-Tree \cite{fischbacher2007z} hosted on our laboratory computers.
At the beginning of each session, all participants  read the instructions presenting the purpose of the experiment, a short description of the four games, the payment procedure and some example screenshots illustrating the interaction of users with the z-Tree tool. Instructions informed participants that they would play the games in an arbitrary order. For each of the games participants were told what partner information would be displayed during each interaction: for the Simple Game no information, for the Identity Game the partner identity in the form of an ID, for the Score Game a partner trust score computed according to her behaviour in previous interactions (without any details about the metric) and for the Combined Game, the partner identity and trust score. Participants did not know the number of rounds they would play in each game. After confirming that they had read and understood the instructions, participants reviewed and signed an informed consent form prior to commencing the experiment. Participants sat in different rooms to avoid any communication during the experiment. Each participant used a computer running our z-Tree application. All senders in the group finished their decision making process before proceeding to the next trial. Play then waited for every receiver to respond before starting a new round.  This eliminated response time cues as an indication of player identity. No other means of communication or identification were available. Participants were informed of their cumulative earnings at each round. It was possible to play with a negative balance but this never occurred.

The repeated measures design resulted in 100 rounds across the four games. A session usually lasted two hours. At the end of the experiment participants filled out a questionnaire regarding general information such as university major and game preference.

\section{Results}

We organize our results into two main subsections:  sender behavior and receiver behavior.  

\subsection {Sender Behavior}

The following analyses address how the sender (trustor) responded to our manipulations.  We demonstrate that both trust score and ID increase sending generosity with equivalent improvement and no combined effect. To examine cooperation, we study the 0 exchange condition and rule-out round effects as influential for all games except the Simple Game with no partner information. Finally, we illustrate the dependence of performance on trust score metrics. 

\begin{figure}[tb]
    \centering
    \includegraphics[width=.5\textwidth]{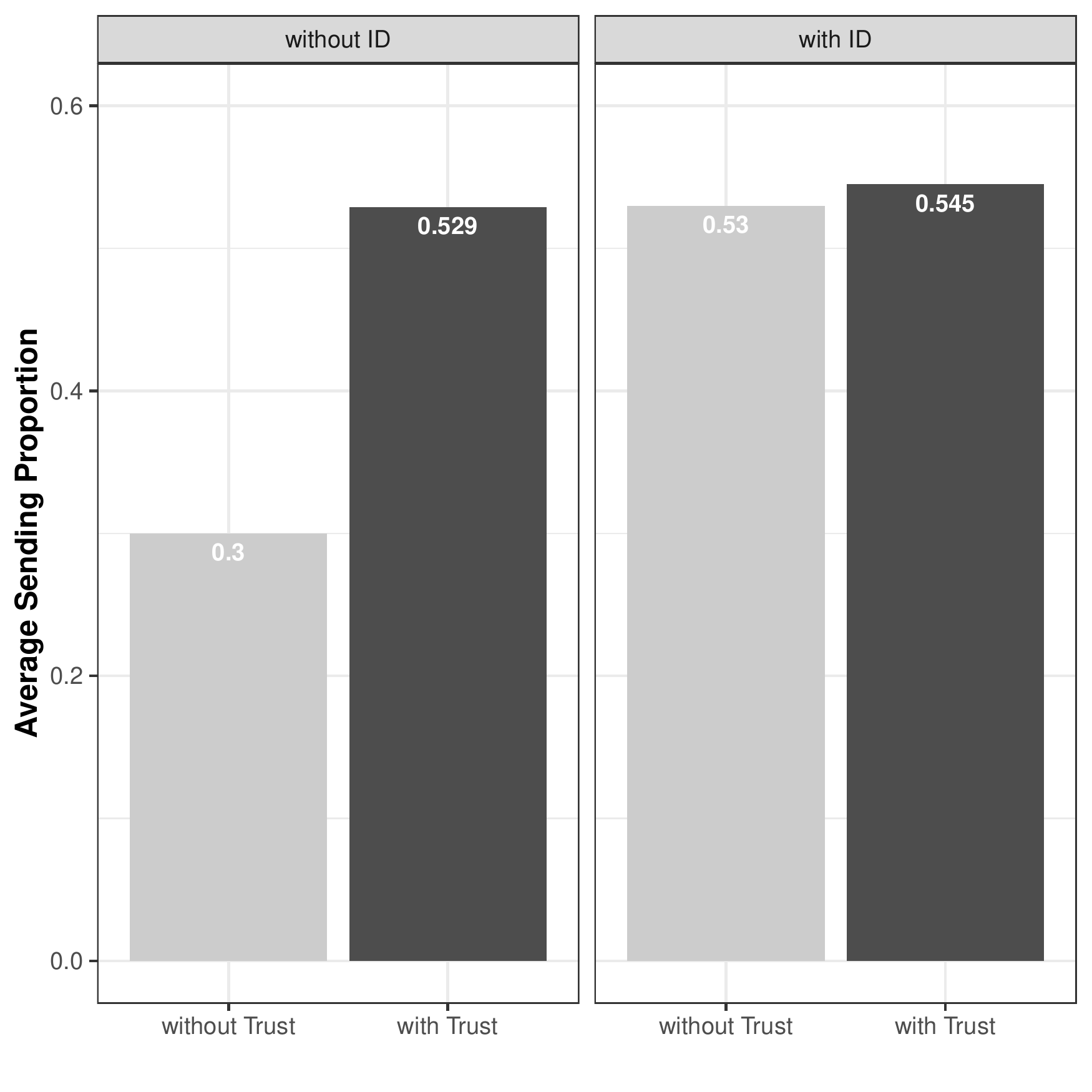}
    \caption{Interaction between trust score and ID availability for sender.}
    \label{fig:send_proportion_sender_by_game}
\end{figure}

\begin{table}[b]
    \small
    \centering
    \begin{tabular}{@{}lll@{}}
        \toprule
        \bf Game in Comparison with Simple Game & \bf 95\% confidence interval& \bf Df  \\
        \midrule
        Identity Game & (-0.32, -0.14) &  29  \\
        Score Game & (-0.35, -0.13) &  29  \\
        Combined Game & (-0.35, -0.14) &  29  \\
        \bottomrule
    \end{tabular}
    \caption{Paired t-based confidence intervals for senders' average sending proportion in the Simple Game compared to other games. The negative signs indicate that the sending amount of participants in Simple Game is less than the sending amount of these participants in other games.}
    \label{tab:t_test_sender_simple_percentage}
\end{table}

\subsubsection{Omnibus ANOVA} A basic ANOVA with Subject, Show-Trust and Show-ID as predictors reveals an interaction,
\emph{F}(1,29) = 19.36, $\emph{p} < 0.001$ as measured by average proportion sent  for each game. The interaction between the availability of trust score and ID on average sending proportion for senders appears in Figure \ref{fig:send_proportion_sender_by_game}. We note that showing either trust score or ID improves sending proportion but showing both partner information sources does not change the sent proportion relative to one source, which suggests the need for paired comparisons between games. \citeN{johnson2011trust} claimed that in large-scale the send proportion of users in trust game follows the normal distribution. Therefore we present paired t-test based confidence intervals (yoking by sender ID)  in Table \ref{tab:t_test_sender_simple_percentage} to examine differences between the Simple Game and any other tested game, demonstrating that either trust score or ID increases sending amounts with no additive effect. The differences between the other three games (Identity, Score and Combined Games) are not significant, i.e. $\emph{p} > 0.10$.  To rule out any possible difference between sender performance with Show-ID and Show-Trust, we followed up with a paired t-test, yoking the results from the Identity Game and the Score Game for each sender-receiver pair for each trial \emph{t}(266) = -0.175, $\emph{p} > 0.10$. We conclude 
$  Identity Game \approx Score Game \approx Combined Game > Simple Game$ for average sending proportion.

\subsubsection{Cooperative Behavior}

Below we address the claim that providing identification or trust score controls cooperative behavior, explaining the above results.  We consider the cases of non cooperation where senders send $0$, the change in trust scores over time and the dependence of sending behavior on trust score values.  

The percentage of times that a sender sends $0$ in Simple Game, Identity Game, Score Game and Combined Game are $33.3\%$, $9.3\%$, $13.6\%$ and $12.7\%$ respectively.  A logistic regression on the frequency of $0$ transactions for all rounds with sending participant, Show-Trust and Show-ID as predictors indicates an interaction between Show-Trust and Show-ID  \emph{z} = 5.607, $\emph{p} < 0.001$. Senders are more likely to send $0$  in the Simple Game.

To examine the potential change in sending behavior over round, we regressed sending behavior on participant ID to remove general participant effects that would contaminate a regression analysis. We then used the resulting residuals as the criterion in a regression with round number as the predictor, reducing the df in the error term due to the prior regression. The only game with a significant round effect was the game with no information (Simple Game), revealing decreasing cooperation over time \emph{F}(1,116) = 7.3, $\emph{p} < 0.01$. No other game indicated a round effect: Identity Game,  \emph{F}(1,114) = 0.05, $\emph{p} > 0.10$, Score Game \emph{F}(1,115) = 0.42, $\emph{p} > 0.10$ and Combined Game  \emph{F}(1,116) = 0.008, $\emph{p} > 0.10$. Partner information eliminates decreasing cooperation over time and end game effects for senders.

Finally, in Table \ref{tab:trust_sender} we present regression analyses between average sending behavior as the criterion with sender trust values and participant trust values as predictors.  Sender behavior is positively correlated with his own trust value for all games. The trust function predicts sender behavior well. Moreover, when partner trust is available, it controls sending behavior. Notably, this is the only analysis suggesting any difference between the availability of partner identity and the trust score, as partner trust score does not predict sending behavior in games without a trust score. We conclude that partner trust score availability controls cooperation. We also note the relatively high adjusted $R^2$ for the Simple Game.  We attribute this to range restriction on trust score values that eliminates non-linear influences at higher levels of trust. 
\begin{table*}[t]
\small
\centering
\begin{tabularx}{\textwidth}{@{}l *{7}{Y}@{}}
\toprule
   & \multicolumn{2}{c}{\bf without Trust} &\multicolumn{2}{c}{\bf with Trust} \\
& {\bf without ID}\,(Simple) & {\bf with ID}\,(Identity) & {\bf without ID}\,(Score) & {\bf with ID}\,(Combine)\\
\midrule
Own trust & 12.80*** & 9.31*** &7.36*** & 8.33***\\

Partner trust  &  1.65 & 1.73 & 5.69*** & 4.69*** \\
Adjusted $R^2$  & 0.85 & 0.75 & 0.88 & 0.89 \\
F(2,27) & 86.03 & 43.57 & 106.9 & 117.1 \\

\bottomrule
\end{tabularx}
\caption{Trust regression analysis for average sending behavior of senders. The table reports on t(27) values. `*'$\emph{p} < 0.05$,  `**' $\emph{p} < 0.01$, `***' $\emph{p} < 0.001$.}
\label{tab:trust_sender}
\end{table*}

\subsubsection{Summary of Sender Behavior} Senders are less cooperative in the Simple Game than all other games.  Decreasing cooperation in the form of round effects only appears in the Simple Game.  Good models for sending behavior show predictive effects of own trust in all conditions, and partner trust when trust scores are available. The availability of partner trust score therefore controls sending behavior. 

\subsection{Receiver Behavior}

The following analyses address how the receiver (trustee) responded to our manipulations. We demonstrate that both trust score and ID increase generosity with equivalent improvement and no combined effect.  To examine cooperation, we study the 0 exchange condition when the receiver received a positive amount from the sender but decided to send back 0. We rule-out round effects and examine the dependence of performance on trust score metrics. 

\begin{table}[b]
    \small
    \centering
    \begin{tabular}{@{}lll@{}}
        \toprule
        \bf Game in Comparison with Simple Game & \bf 95\% confidence interval& \bf Df  \\
        \midrule
        Identity Game & (-0.23, -0.10) &  29  \\
        Score Game & (-0.25, -0.08) &  29  \\
        Combine Game & (-0.26, -0.11) &  29  \\
        \hline
    \end{tabular}
    \caption{Paired t-test confidence intervals for receivers' average sending proportion in Simple Game compared to other games. The negative signs indicate that the sending amount of participants in Simple Game is less than the sending amount of these participants in other games.}
    \label{tab:t_test_receiver_simple_percentage}
\end{table} 

\subsubsection{Omnibus ANOVA}
\begin{figure}[t]
    \centering
    \includegraphics[width=.5\textwidth]{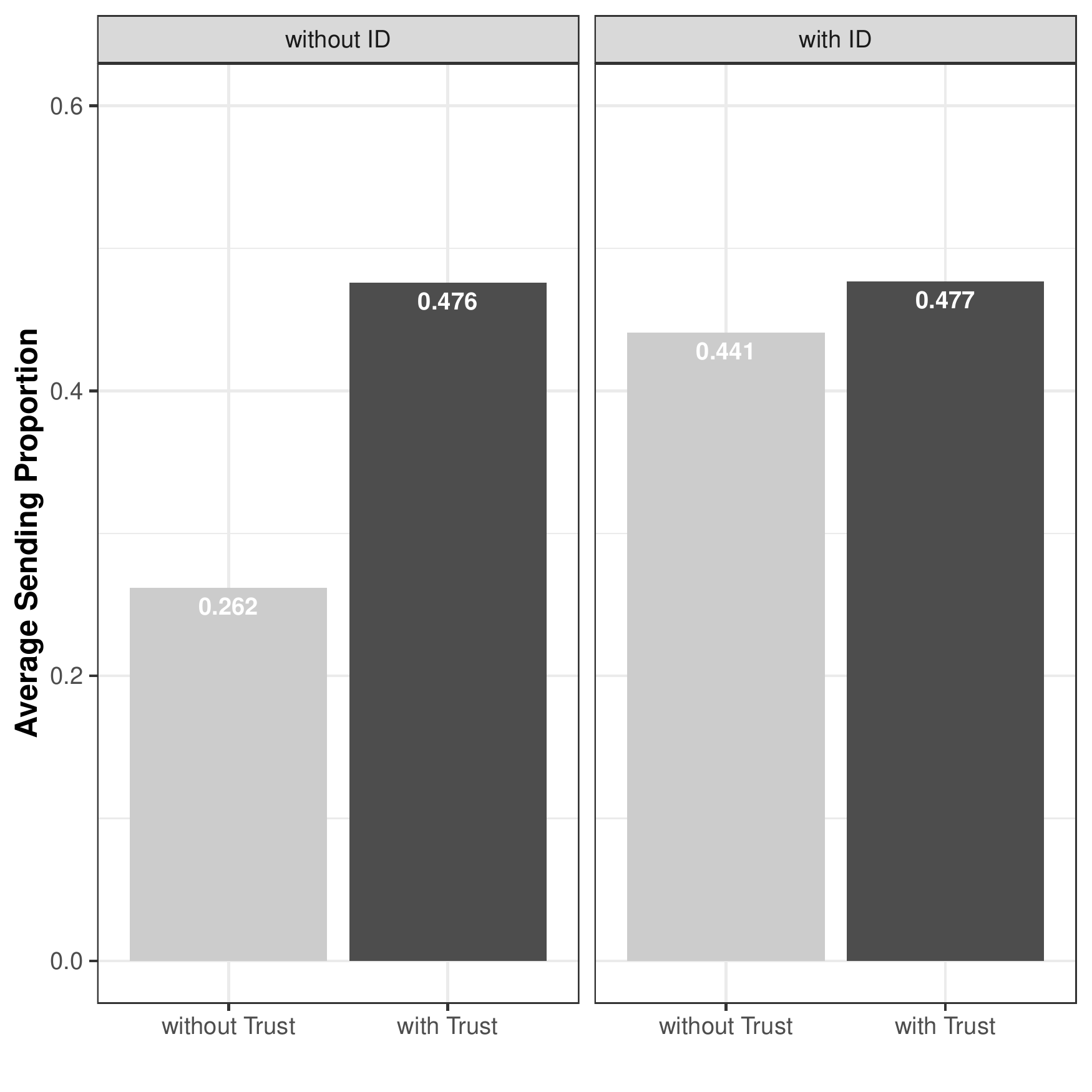}
    \caption{Interaction between trust score and ID availability for receiver.}
    \label{fig:send_proportion_receiver_by_game}
\end{figure}

A basic ANOVA with Subject, Show-Trust and Show-ID as predictors reveals an interaction,
\emph{F}(1,29) = 14.36, $\emph{p} < 0.001$ as measured by average sending proportion.  The interaction between the availability of trust score and ID on average sending proportion appears in Figure \ref{fig:send_proportion_receiver_by_game}. We note that showing either trust score or ID improves receiver return proportions, but showing both partner information sources does not change the sent amount relative to one source which suggests the need for paired comparisons between games. As above, and consistent with \citeN{johnson2011trust} we assume that the sending proportion of receivers follows the normal distribution in large-scale. We used paired-t based confidence intervals (yoking by receiver ID) in Table  
\ref{tab:t_test_receiver_simple_percentage} to examine differences between the Simple Game and any other tested game. Showing either trust score or ID increases the amount sent back with no additive effect. To rule out any possible difference between receiver performance with Show-ID and Show-Trust, we followed up with a paired t-test yoking the results from the Identity Game and the Score Game for each receiver-sender pair for each trial. The results of the paired t-test, i.e. \emph{t}(219) = -0.458, $\emph{p} > 0.10$ confirmed the absence of difference between  Identity Game and  Score Game. We conclude 
$Combined Game \approx Score Game \approx Identity Game > Simple Game$ for receiving behavior. 

\subsubsection{Cooperative Behavior}
Below we address the claim that providing identification or trust score increases cooperative behavior, explaining the above results.  We consider the cases of sending 0, the change in trust scores over time and the dependence of receiver behavior on trust score values.  

The percentage of times that a receiver sends $0$ in Simple Game, Identity Game, Score Game and Combine Game are $36.8\%$, $8.5\%$, $8.3\%$ and $4.5\%$ respectively.  A logistic regression on the frequency of $0$ transactions for all trials with sending participant, Show-Trust and Show-ID as predictors indicates an interaction between Show-Trust and Show-ID  \emph{z} = 3.68, $\emph{p} < 0.01$. Receivers are more likely to return 0  in the Simple Game.

To examine the potential change in receiver behavior over round, we regressed receiver behavior on participant ID to remove general participant effects that would contaminate a regression analysis. We then used the resulting residuals as the criterion in a regression with round number as the predictor, reducing the df in the error term due to the prior regression. Round is not significant for any game: Simple Game \emph{F}(1,100) = 0.052, $\emph{p} > 0.10$, Identity Game,  \emph{F}(1,114) = 1.44, $\emph{p} > 0.10$, Score Game \emph{F}(1,108) = 0.019, $\emph{p} > 0.10$ and Combined Game \emph{F}(1,110) = 0.027, $\emph{p} > 0.10$.  Participant information therefore has no effect on the prevention of end-game effects, which do not exist.

Finally, in Table \ref{tab:trust_receiver} we present regression analyses between average sending behavior as the criterion with sender trust values, participant trust values and amount received from the sender as predictors.  Receiver behavior is positively correlated with his own trust value for all games. This confirms our ability to predict receiver cooperation (i.e., receiver trustworthiness) from past trust values. However, receiver behavior is only related to partner trust in the Combined Game. Moreover, model fits are not as good for receivers as they are for senders. We have explored models that include interactions between amount received and trust values.  These often improve the relatively smaller adjusted $R^2$ we obtain for receiver behavior. Such models suggest the need for different trust functions for sender and receiver, to accommodate the asymmetry in their relationship.

\begin{table*}[t]
\small
\centering
\begin{tabularx}{\textwidth}{@{}l *{7}{Y}@{}}
\toprule
   & \multicolumn{2}{c}{\bf without Trust} &\multicolumn{2}{c}{\bf with Trust} \\
& {\bf no ID}\,(Simple) & {\bf with\,ID}\,(Identity) & {\bf no ID}\,(Score) & {\bf with\,ID}\,(Combined)\\
\midrule
Own trust & 6.003*** & 8.936*** &4.617*** & 3.927***\\

Partner trust  &  0.687 & 0.978 & 0.237 & -2.158* \\
Partner sending amount  & -2.214* & -1.849 & -1.469 & 0.587 \\
Adjusted $R^2$  & 0.565 & 0.746 & 0.415 & 0.494 \\
F(3,26) & 13.53 & 29.36 & 7.854 & 10.44 \\

\bottomrule
\end{tabularx}
\caption{Trust regression analysis for average sending behavior of receivers. The table reports on t(26) values.`*'$\emph{p} < 0.05$,  `**' $\emph{p} < 0.01$, `***' $\emph{p} < 0.001$.}
\label{tab:trust_receiver}
\end{table*}

\subsubsection{Summary of Receiver Behavior} Receivers are less cooperative in the Simple Game than all other games.  There is no evidence of round effects in any game.  Fair models for returning behavior show predictive effects of own trust in all conditions confirming our trustworthiness predictions. However, partner trust is only predictive in the combined game.

\section{Experimental Design Issues} 
In this section we investigate the properties of our experiment, comparing our results with other trust game experiments, evaluating the accuracy of our trust function, and addressing repeated measures concerns such as the nesting of participants in groups.

\subsection{Comparison with other trust game data sets}

\begin{figure}[h]
\centering
\includegraphics[width=0.5\textwidth,trim={0 1.5cm 0 2cm},clip]{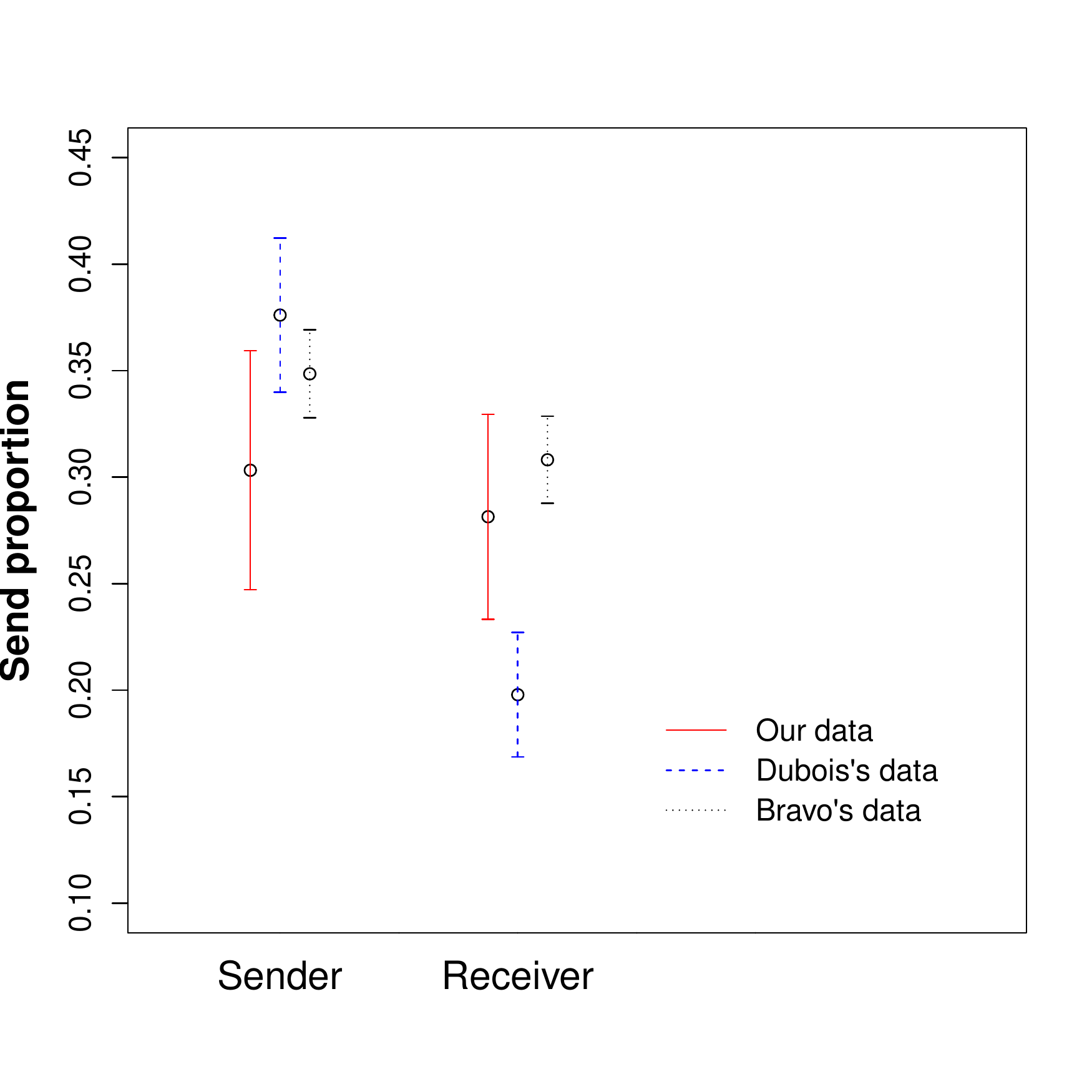}
\caption{Average values and standard errors of users' sending proportions in three datasets.}
\label{fig:compare_3_data}
\end{figure}

We compared the average sending proportions of participants in our Simple Game (30 data points) with two external datasets from \citeN{dubois2012does} with 36 data points and \citeN{bravo2012trust} with 108 data points. Table  \ref{tab:compare_send_3_data} shows Welch two-sample t-test values comparing our results in the simple game to their results, assuming unequal variances. None of the comparisons are statistically significant.  The observed behavior in the simple game in our experimental design is consistent with other experiments. The findings are illustrated in Figure \ref{fig:compare_3_data}.

\begin{table}[b]
\small
\centering
\begin{tabular}{@{}lrr@{}}
    \toprule
    & \bf \citeN{dubois2012does} & \bf \citeN{bravo2012trust} \\
    \midrule
    Sender & t(61.6) = -1.33 & t(45.3) = -0.991 \\
    Receiver & t(55.9) = 1.69 & t(45.6) = -0.598 \\
    \bottomrule
\end{tabular}
\caption{Welch two-sample $t$ values between our Simple Game average send proportion data with two external datasets.}
\label{tab:compare_send_3_data}
\end{table}

\subsection{Trust function analysis}

In the previous sections, we demonstrated that showing the trust score improves cooperation, but how good is the trust function? We provide two forms of support for the quality of the trust function: detailed prediction of participant behavior in our experiment and prediction of participant behavior using calculated reputation.

\subsubsection{Predicting behavior in our experiment.}
The trust score models participant behavior, even when, as in Simple and Identity Games, the trust score is not made available to participants. Thus participant behavior should correlate with their own trust scores \cite{DBLP:conf/trustcom/DangI16}. In the games with presented trust scores (Score and Combine Games), participant behavior should appear to react to partner trust values. The $R^2$ values in Tables \ref{tab:trust_sender} and \ref{tab:trust_receiver} provide some evidence of prediction accuracy, although we noted less satisfactory models for receivers, and less evidence for the relevance of partner trust values in receiver behavior.  Here we rule out interactions between trust values themselves as better behavior predictors. We also examine correlations between behavior and trust scores separately for rounds 4 and 5 when trust scores have sufficient data to stabilize. 

Regressions of sender behavior, i.e. average sending proportion, on the interaction of sender and receiver trust values in the presence of both predictors as main effects provide no evidence of interaction effects in any game:   Score Game \emph{t}(26) = 1.079, $\emph{p} > 0.1$, Combined Game \emph{t}(26) = 0.022, $\emph{p} > 0.1$, Simple Game \emph{t}(26) = -0.352, $\emph{p} > 0.1$  nor Identity Game \emph{t}(26) = 0.725, $\emph{p} > 0.1$.

Regressions of receiver behavior, i.e., average return proportion, on the interaction of sender and receiver trust values in the presence of both predictors as main effects provide no evidence of interaction effects in any game:   Score Game \emph{t}(26) = -0.122, $\emph{p} > 0.1$, Combined Game \emph{t}(26) = -0.776, $\emph{p} > 0.1$, Simple Game \emph{t}(26) = 0.706, $\emph{p} > 0.1$  nor Combined Game \emph{t}(26) = 0.080, $\emph{p} > 0.1$. Adding interactions between trust predictors does not improve our models.

To further examine the predictive capability of the trust function, we performed separate multiple regression analyses for each game, for rounds 4 and 5 when trust scores have accrued sufficient data. The criterion variable is the sending proportion of the participants to their partners. Table \ref{tab:regression_sender} provides the results of a regression of the senders sending proportion on a model with her trust value and the trust value of her partner for both rounds.  In all cases, the sender's trust value predicts sending behavior.  Moreover, the partner's trust value also predicts sending behavior in the presence of ID or trust score information, confirming sender attention to these sources. Adjusted $R^{2}$ values range from 0.26 to 0.70, with lower values resulting from the game with no information. 

\begin{table*}[h]
\small
\centering
\begin{tabularx}{\textwidth}{@{}l *{7}{Y}@{}}
\toprule
   & \multicolumn{2}{c}{\bf without Trust} &\multicolumn{2}{c}{\bf with Trust} \\
& {\bf no ID}\,(Simple) & {\bf with\,ID}\,(Identity) & {\bf no ID}\,(Score) & {\bf with\,ID}\,(Combine)\\
\midrule
Round 4 & df = 72 & df = 72&df = 72& df = 72\\
Own trust value  &  6.46*** & 5.80*** & 3.89*** & 7.28*** \\
Partner's trust value  & 0.67 & 3.24** & 6.98*** & 4.41*** \\
Adj. $R^2$ & 0.36*** & 0.40*** & 0.66*** & 0.70*** \\
\\[-.5em]
Round 5 & df = 72 & df = 72 & df = 72 & df = 72\\
Own trust value  & 4.87*** & 7.13*** & 3.19** & 7.11*** \\
Partner's trust value  & 1.16 & 4.54*** & 7.38*** & 3.52*** \\
Adj. $R^2$ & 0.26*** & 0.55*** & 0.67*** & 0.70*** \\
\bottomrule
\end{tabularx}
\caption{Trust regression analysis on senders' sending proportion with t-values for individual slope tests. `*'$\emph{p} < 0.05$,  `**' $\emph{p} < 0.01$, `***' $\emph{p} < 0.001$.}
\label{tab:regression_sender}
\end{table*}%

Table \ref{tab:regression_receiver} provides comparable information for receiver behavior, answering the question of how well we can predict whether a participant is trustworthy. These regression models included own trust value, partner trust value and the amount just received (i.e., three times the amount sent). While receivers were never aware of their own trust values, our trust function is a good predictor of receiver behavior \emph{when trust score is not provided.} This does support our claim that the trust function is a good predictor of trustworthiness.  However, the mere presence of trust scores in the trust score conditions dampens its predictive capability.  Partner trust value is rarely predictive. Receivers did not rely on this systematically. Adjusted $R^{2}$values range from 0.08 to 0.45 with higher values in the conditions where trust score is \emph{not} provided.

\begin{table*}[b]%
\small
\centering
\begin{tabularx}{\textwidth}{@{}l *{7}{Y}@{}}
\toprule
   & \multicolumn{2}{c}{\bf without Trust} &\multicolumn{2}{c}{\bf with Trust} \\

& {\bf no ID}\,(Simple) & {\bf with\,ID}\,(Identity) & {\bf no ID}\,(Score) & {\bf with\,ID}\,(Combine)\\
\midrule
Round 4 & df = 42 & df = 62 & df = 60 & df = 60\\
Own trust value  &  3.41** & 7.21*** & 1.98 & 1.76 \\
Partner's trust value  & 0.02 & 1.40 & 1.63 & 0.50 \\
Amount received & -0.53 & -1.62 & -2.37*& 0.33\\
Adj. $R^2$ & 0.18* & 0.45*** & 0.08 & 0.10* \\
\\[-.5em]
Round 5 & df = 39 & df = 61 & df = 61& df = 60\\
Own trust value  & 4.21*** & 3.56*** & 3.06** & 1.09 \\
Partner's trust value  & 0.14 & 2.10* & 0.74 & 1.53 \\
Amount received &-2.19* & 0.06& -1.75 & -0.16\\
Adj. $R^2$ & 0.30*** & 0.29*** & 0.13* & 0.09* \\
\bottomrule
\end{tabularx}
\caption{Trust regression analysis on receivers' sending proportion with t-values for individual slope tests. `*'$\emph{p} < 0.05$,  `**' $\emph{p} < 0.01$, `***' $\emph{p} < 0.001$.}
\label{tab:regression_receiver}
\end{table*}%

\subsubsection{Post-hoc Reputation Analysis}

We present a post-hoc analysis to compare the predictive power of participants future behavior  between trust and reputation scores.

In our analyses presented in Tables \ref{tab:trust_reputation_sender_average} and \ref{tab:trust_reputation_receiver_average} we
substituted calculated reputation predictors for trust predictors, using average sending proportion as the criterion. These models differ from those in Tables \ref{tab:trust_sender} and \ref{tab:trust_receiver} by the absence of own-score predictors. These reduced models were necessary because of the close relationship between average reputation and average sending amount.  However, the absence of own-values does inflate the error term.  As in Table \ref{tab:trust_sender}, in Table \ref{tab:trust_reputation_sender_average}  partner reputation values predict sender behavior when trust values are shown. As measured by Adjusted $R^2$, the resulting models of sender behavior with trust predictors are better than models with reputation predictors. This is not surprising given that we provided trust values and not reputation values in these conditions.  However, the superior fit provides further evidence that participants were attending to the trust values.  Regarding receiver behavior, in Table \ref{tab:trust_receiver} partner trust is only significant in the Combined game. In Table \ref{tab:trust_reputation_receiver_average} partner reputation predicts receiver behavior for the ID game, no doubt assisted by the significant effect of partner sending amount. We note that in those cases with significant partner effects, the direction is negative regarding to the amount received.  Model fits are poor. Adjusted $R^2$ are, however, better for trust predictors  than  reputation predictors for the games where trust information was present.

\begin{table*}[t]%
\small
\centering
\begin{tabularx}{\textwidth}{@{}l *{7}{Y}@{}}
\toprule
  & \multicolumn{2}{c}{\bf without Trust} &\multicolumn{2}{c}{\bf with Trust} \\
& {\bf no ID}\,(Simple) & {\bf with\,ID}\,(Identity) & {\bf no ID}\,(Score) & {\bf with\,ID}\,(Combine)\\
\midrule
\bf Trust predictors \\
Partner trust  &  1.09 & 0.33 & 7.42*** & 6.92*** \\
Adjusted $R^2$  & 0.007 & -0.03 & 0.65 & 0.62 \\
F(1,28) & 1.202 & 0.11 & 55.07*** & 47.86*** \\
\\[-.5em]
\bf Reputation predictors \\
Partner reputation  &  0.69 & -1.14 & 4.55*** & 3.78*** \\
Adjusted $R^2$  & -0.01 & 0.01 & 0.40 & 0.31 \\
F(1,28) & 0.48 & 1.3 & 20.72*** & 14.31*** \\
\bottomrule
\end{tabularx}
\caption{Trust and reputation analysis for average sending proportion of senders. The table reports on t(28) values. `*'$\emph{p} < .05$,  `**' $\emph{p} < .01$, `***' $\emph{p} < .001$.}
\label{tab:trust_reputation_sender_average}
\end{table*}%

\begin{table*}[b]%
\small
\centering
\begin{tabularx}{\textwidth}{@{}l *{7}{Y}@{}}
\toprule
   & \multicolumn{2}{c}{\bf without Trust} &\multicolumn{2}{c}{\bf with Trust} \\
& {\bf no ID}\,(Simple) & {\bf with\,ID}\,(Identity) & {\bf no ID}\,(Score) & {\bf with\,ID}\,(Combine)\\
\midrule
\bf Trust predictors \\
Partner trust  &  -0.71 & -0.41 & -0.26 & -2.73* \\
Partner sending amount  & -0.22 & 1.35 & 0.85 & 3.20* \\
Adjusted $R^2$  & 0.00 & 0.00 & -0.02 & 0.22 \\
F(2,27) & 0.99 & 1.05 & 0.64 & 5.18* \\
\\[-.5em]
\bf Reputation predictors \\
Partner reputation  &  -1.70 & -2.72* & -0.15 & -1.40 \\
Partner sending amount  & 0.23 & 2.33* & 0.45 & 2.07* \\
Adjusted $R^2$  & 0.08 & 0.21 & -0.02 & 0.08 \\
F(2,27) & 2.26 & 4.93 & 0.62 & 2.21 \\
\bottomrule
\end{tabularx}
\caption{Trust and reputation analysis for average sending proportion of receivers. The table reports on t(27) values. `*'$\emph{p} < .05$,  `**' $\emph{p} < .01$, `***' $\emph{p} < .001$.}
\label{tab:trust_reputation_receiver_average}
\end{table*}%

\subsection{Group Effects}  While data on the trust game are typically collected in groups, concern for group effects has received little attention in trust game analyses. Moreover, in our experiment, group is confounded with treatment order.  In order to consider group effects, we conducted a three factor split-plot ANOVA with group as a between subjects effect and Show-ID and Show-Trust as within subjects effects \cite{keppel1991design}.  If group is regarded as a random (sampled) factor, then the independent variables are properly tested against the interaction of group with the independent variables.  

Our sole concern here therefore is the robustness of manipulation effects in a very conservative, low power test owing to the reduced df in the error term. We tested our effects considering group as a random factor, and interactions with group as an error term.  Our analysis of sending behavior, as measured by relative sending proportion, withstands even this less powerful test. The omnibus test for the interaction of ID and Trust is \emph{F}(1,4) = 8.86, $\emph{p} < 0.05$.  Moreover, none of the Group by Treatment interactions are significant: with Show-Trust \emph{F}(4,25) = 2.610, $\emph{p} > 0.05$, with Show-ID \emph{F}(4,25) = 1.253, $\emph{p} > 0.05$, or the interaction \emph{F}(4,25) = 2.698, $\emph{p} > 0.05$.
Regarding receiver behavior, as measured by relative returned proportion, the omnibus interaction contrast just misses significance  \emph{F}(1,4) = 6.966, $\emph{p} < 0.1$.  These findings are best captured as two main effects:  for Show-Trust \emph{F}(1,4) = 74.44, $\emph{p} < 0.001$ and for Show-ID \emph{F}(1,4) = 35.862, $\emph{p} < 0.01$). As above, none of the Group by treatment interactions are significant:  with Show-Trust \emph{F}(4,25) = 0.153, $\emph{p} > 0.75$, with Show-ID \emph{F}(4,25) = 0.553, $\emph{p} > 0.75$, or the interaction \emph{F}(4,25) = 2.484, $\emph{p} > 0.05$.
These analyses limit concern for group effects in general, and the game order differences confounded with group in particular.  

\label{subsec:group_effect_game_info}.

\section{Discussion} 
 We analyzed our research questions distinguishing between sender's and receiver's points of view.

\textbf{RQ1} \textit{Does showing partner trust score or ID change user cooperative behavior?}

We provided several forms of evidence regarding the influence of these interventions on cooperation.  These include overall increases in the proportion returned and reductions in the frequency of 0 unit returns for both senders and receivers. Only the Simple Game differs from the alternatives, in paired-t tests of sending behavior and in the persistence of end-game effects for senders. Otherwise, we eliminated end game effects. Large-n, yoked dependent t-tests by round failed to reveal any difference in behavior between the availability of names and the availability of trust scores.

\textbf{RQ2} \textit{Does the trust calculation predict participants' future behavior ?}

With respect to senders, we provide excellent predictive models for average behavior.  These average models always depend positively on own trust values, and on partner trust values when trust values are available. Sender behavior is also well modeled at the round level, always depending upon own trust values and on partner trust values for all games except the Simple Game. Senders are attending to the specific values shown for partners, as predictions based on reputation are not as good as predictions based on the trust values displayed. We note that the effect is not to encourage blind cooperation, but rather cooperation in response to the available information.  Low partner trust scores elicit low sending amounts.   

With respect to receivers, models of average return proportions behavior do depend on own-trust.  This supports a claim for some ability to predict trustworthiness. Models at the round level are best when the trust score is \emph{not} available. This unexpected result is possibly due to strategic differences in receiver behavior. Models are quite poor when own-values are removed in order to compare with reputation predictions. While receiver models did include an additional factor (partner sending amount), our general impression is that the models of receiver behavior are more complex than models of sender behavior and not yet accommodated by the trust function used. Moreover, unlike the sender, duplicitous receiver behavior is not punished until the subsequent round. These considerations suggest that the trust function should differ for sender and receiver.   

We have not identified the source of leverage on the success of the trust function for senders. Relative to an average reputation calculation, we have noted three different influences:  the specification of partners, the management of change over time and the treatment of variability, particularly punishment in response to non-cooperative behavior. Limitations in the receiver model highlight this claim, where the role of amount received may interact with the partner trust values in ways that we have not yet captured. These influences cast the trust function as a psychometric issue, concerning the psychological factors that influence the response to experience. 

Our preliminary study suggested the predictive power of trust scores compared to that of reputation. However, our experimental study did not include a condition with computed reputation score for the partner in order to be able to compare the influence of trust score and reputation.

The trust function used considers only the sending proportion as a parameter, but not for instance the amount sent by the partner. This trust model fits well for a sender that initiates the interaction by sending an initial amount. But the trustworthiness value associated with a receiver should depend not only on the return proportion but also on the amount received. We might consider associating a higher trustworthiness with a receiver that received $6$ and returned $1$ than to someone that received $30$, but returned the same proportion. The receiver that received $30$ obtained the maximum possible amount but did not reciprocate the granted trust. These suggestions further reinforce the need to consider the measurement of trust from a psychometric perspective, capturing the relationship between physical quantities and behavioral response.

We have demonstrated that the presence of partner information benefits cooperative behavior.  The burden of recalling past experience with participants is just one justification for the use of trust values as a source of this information \cite{tang2013social}. 
Compared with reputation scores, trust scores have several advantages. Reputation scores are globally computed values that are stored on a central server that is vulnerable to attack \cite{DBLP:journals/csur/HoffmanZN09}. Trust scores are suitable for \textit{distributed} architectures and do not require a central server.
Trust scores are computed in a distributed way for each user: each member of the network locally computes trust levels of her partners. Moreover, trust scores emphasize \textit{personal} experience and value. For instance, in reputation systems, if ten thousand participants rated a seller, the next participant does not have a high motivation to provide a rating because it will not change the average rating score of this seller. However, in trust-based systems, her impression has a great influence because the trust value is calculated for her only based on her experience. 

On the other hand, as our experiment suggested, the trust score has a similar effect on cooperative behavior relative to ID. Therefore, trust scores may complement current systems that employ ID to identify users, helping users define the trustworthiness of their connections. While it is possible for participants to change their ID in on line systems, they cannot change the trust level other participants assigned to them. If a trust score is available, participants do not need to remember individuals by name, nor do they need to assess previous experience with imprecise mental calculations.  Instead, they can make decisions based on their partner's current trust score. 

Such a system greatly facilitates engagement with large scale collaborative networks. Our proposed solution for computing partner trust scores scales well with the number of partners. For each user $u_i$, where $1 \leq i \leq n$ and $n$ is the total number of partners, the system stores $m_i$ trust values $t_{ij}$, with $1 \leq j \leq m_i$, associated with the $m_i$ partners with whom he is interacting. Each time a participant $u_i$ interacts  with another partner $u_j$, the trust score corresponding to that interaction is aggregated to the old trust value $t_{ij}$. The new aggregated value becomes the new value of $t_{ij}$. The time complexity of the computation of the trust score from an interaction is $O(1)$, i.e. constant. The space complexity for a participant to keep track of the trust scores of the other participants is linear with the number of participants with whom he interacts.

Regarding generalizability, significant effort remains in developing trust functions for other domains.  Our claim is not that the specific function we used \cite{DBLP:conf/trustcom/DangI16} is suitable for every domain, but rather that the dimensions we have identified (partner specificity, the representation of cumulative experience, and the treatment of variability) are candidates for inclusion. 

\section{Conclusions} \label{sec:conclusion}

We showed that trust score or ID availability could significantly improve the level of cooperation between users. We also demonstrated that the availability of a trust score has a similar impact on boosting cooperation as the availability of identities. Finally we showed that the availability of both features has no additional benefit to cooperation as the availability of only one of these features. Our study suggests that trust score could function as an enhancement or even replacement of traditional ID systems. We plan to study a closer comparison between the influence of trust score and reputation score on the collaborative behavior by designing a trust game experiment where we analyse the effect of showing partner trust score and reputation score.

\section{Acknowledgments}
This work was partially supported by Inria Associated Team USCoast and NSF Grant 1520870.

\bibliographystyle{ACM-Reference-Format}
\bibliography{references}

\appendix
\section{Trust Score Calculation} \label{ap:trust_calc}

Separate trust scores are calculated for each player for each round, i.e. for each interaction between two players. The round number is denoted as $t$. 

In our experiment, a user might be assigned different role in different round, i.e. in a round she can be a sender and in another round she can be a receiver. The maximum amounts she can send are different by role, which is $10$ in case of sender and $3*received\_amount$ in case of receiver. Therefore, we firstly normalize the sending amount of both roles to $send\_proportion_t$ for round $t$.

\begin{equation} \label{eq:send_proportion}
send\_proportion_t = \frac {\sendingamount{t}}{\maxsendingamount{t}}
\end{equation}

The zero-transaction is eliminated on the receiver's side, i.e. if the sender sends 0 to the receiver the trust score of the receiver is kept the same for the next interaction because her send proportion is 0/0, which is undefined. In this case, for this round, the trust score of the sender is updated to 0, send proportion being 0/10=0.0.

Then we calculate the trust score for a single current round $t$:

\begin{equation} \label{eq:current_trust}
\currenttrust{t} = \log (\sendproportion{t} \times (e-1) + 1)
\end{equation}

Then we calculate the aggregate trust score, which is the cumulative trust score over multiple interactions.

\begin{align} \label{eq:alpha_beta_delta}
\delta_t = {} &\abs{\currenttrust{t} - \currenttrust{t-1}} \\
\beta_t = {} &c \times \delta_t + (1 - c) \times \beta_{t-1} \\
\alpha_t = {} & \threshold{} + \frac{c \times \delta_t}{1 + \beta_t}
\end{align}
\begin{align} \label{eq:aggregate_trust}
\aggregatetrust{t} = {} & \alpha_t \times \currenttrust{t} 
                         + (1 - \alpha_t) \times \aggregatetrust{t-1} 
\end{align}

The $\delta_t$ is the change in current trust value by two sequential interactions $t$ and $t-1$ between two users with $\currenttrust{0}=0$. We calculated $\delta_t$ to see how much a person changes her behavior since her last activity. It is easy to prove that, $\alpha_t$ is bigger if $\delta_t$ is bigger, and vice versa. Therefore, if the trust of the current interaction is much different from accumulated trust of all previous interactions, the current interaction will play a more important role in the final trust value. 

Now we can calculate the $\trendfactor{t}$ at round $t$ representing the recent trend of user behavior, with higher values meaning that users improved lately their behavior.
$\trendfactor{t}$ helps us to deal with \textit{fluctuating} behavior, i.e. a user firstly cooperates to gain trust from partners then suddenly deviates: this kind of behavior will be punished immediately by our trust metric.

$\atf{}$ represents accumulated trust fluctuation. Both kinds of \textit{fluctuating behaviors} are punished: whether the latest sending amount is suddenly higher or lower than usual behavior. However, it is obvious that the latter case is more critical than the former one. Therefore, punishment in the latter case should be greater. The accumulated trust fluctuation is a non-decreasing function. The increase depends on the change over time of the user's behavior. If the behavior is stable or it changes within the allowed range (defined by the constant $\phi$), $\atf{t}$ will not change. When $\atf{t}$ reaches the threshold value $\MAXATF{}$, 
accumulated change in user behavior over time reaches the level of betrayal and therefore $\changerate{t}$ drops to 0. Otherwise,  $\changerate{t}$ decreases if $\atf{t}$ increases. The cosine function is used in the formula as it has a low degradation rate in the initial stage, and a high degradation rate in the case of repeated fluctuating behavior. Therefore,
if a user begins to adopt fluctuating behavior the punishment is small, but it increases quickly when fluctuating behavior persists.

\begin{align} \label{eq:trend_factor}
\trendfactor{t} & = 
&
\begin{cases}
\trendfactor{t-1} + \phi 
 \; \text{ if }  \currenttrust{t} - \aggregatetrust{t} > \epsilon \\
\trendfactor{t-1}-\phi 
 \; \text{ if } \aggregatetrust{t} - \currenttrust{t} > \epsilon \\
\trendfactor{t-1}
 \; \text{  otherwise}
\end{cases} 
\end{align}

\begin{equation} \label{atf2}
\adjustedatf{t} = 
\begin{cases}
\frac{\atf{t}}{2} \; \text{ if } \atf{t} > \MAXATF{} \\
\atf{t} \; \text{ otherwise}
\end{cases}
\end{equation}

\begin{equation} \label{eq:atf1}
\atf{t} = \\
\begin{cases}
\adjustedatf{t-1} + \frac{(\currenttrust{t} - \aggregatetrust{t})}{2} 
\; \text{ if } \currenttrust{t} - \aggregatetrust{t} > \phi\\
\adjustedatf{t-1} + (\aggregatetrust{t} - \currenttrust{t}) 
\; \text{ if } \aggregatetrust{t} - \currenttrust{t} > \phi\\
\adjustedatf{t-1} \; \text{ otherwise}
\end{cases}
\end{equation}

\begin{equation} \label{eq:change_rate}
\changerate{t} = \\
\begin{cases}
0 \; \text{ if } \atf{t} > \MAXATF{}\\
\cos{(\frac{\pi}{2} \times \frac{\atf{t}}{\MAXATF{}})} \; \text{ otherwise} 
\end{cases}
\end{equation}

Finally, the trust score is calculated:
\begin{align} \label{eq:trust}
\trustvalue{t} = {}  & \expecttrust{t} \times \changerate{t} \; \text{where, } \nonumber\\
\expecttrust{t} = {} & \trendfactor{t} \times \currenttrust{t} \nonumber
                      + (1 - \trendfactor{t}) \times \aggregatetrust{t} \nonumber
\end{align}

More details about our trust function and its evaluation can be found in \citeN{DBLP:conf/trustcom/DangI16}.

\end{document}